\documentclass[12pt,onecolumn,twoside,draftcls]{IEEEtran}
\date{}

\usepackage{graphics,epsfig,slashbox}
\usepackage{graphicx}
\usepackage{subfigure}
\usepackage{amssymb, amsmath}
\usepackage{multirow}
\usepackage{algorithm}
\usepackage{algorithmic}
\usepackage{url}
\usepackage{afterpage}
\usepackage{color}
\usepackage{epstopdf}

\newcommand{\ignore}[1]{}

\newcommand{\x}{{\bf x}}

\newcommand{\z}{{\bf z}}
\newcommand{\q}{{\bf q}}

\newcommand{\X}{{\bf {X}}}
\newcommand{\Z}{{\bf {Z}}}
\newcommand{\N}{{\mathbb{N}}}

\newcommand{\Ex}{{\mathbb{E}}}

\newcommand{\prob}{{\mathsf P}}
\newcommand{\argmax}{\operatornamewithlimits{argmax}}

\newcommand{\tends}{\rightarrow}

\newcommand{\qed}{\hspace*{\fill}~{\rule{2mm}{2mm}}\par\endtrivlist\unskip}

\newcommand{\complex}{{\mathbb C}}

\newcommand{\snr}{{\mathsf{SNR}}}

\begin{document}

\title{
\LARGE Phase-Quantized Block Noncoherent Communication}

\author{\large
 Jaspreet Singh
        and~Upamanyu Madhow,~\IEEEmembership{Fellow,~IEEE}
\thanks{Jaspreet Singh is with Samsung Telecommunications America, Richardson, TX 75082, USA. Upamanyu Madhow is with the Department of Electrical
and Computer Engineering, University of California, Santa
Barbara, CA 93106, USA. E-mail: \{jsingh@sta.samsung.com, madhow@ece.ucsb.edu\}.

This research was supported by the National Science Foundation under grants CCF-0729222 and CNS-0832154.

This work was presented in part at the IEEE International Symposium on Information Theory (ISIT 2009, Seoul, Korea) and at the IEEE International Conference on Ultra-WideBand (ICUWB 2009, Vancouver, Canada).
.}}

\maketitle \vspace{-11mm}

\begin{abstract}
Analog-to-digital conversion (ADC) is a key bottleneck in scaling DSP-centric receiver architectures to multiGigabit/s speeds. Recent information-theoretic results, obtained under ideal channel conditions (perfect synchronization, no dispersion),  indicate that low-precision ADC (1-4 bits) could be a suitable choice for designing such high speed systems. In this work, we study the impact of employing low-precision ADC in a {\it carrier asynchronous} system. Specifically, we consider transmission over the block noncoherent Additive White Gaussian Noise (AWGN) channel, and investigate the achievable performance under low-precision output quantization. We focus attention on an architecture in which the receiver quantizes {\it only the phase} of the received signal: this has the advantage of being implementable without automatic gain control, using multiple 1-bit ADCs preceded by analog multipliers. For standard uniform Phase Shift Keying (PSK) modulation, we study the structure of the transition density of the resulting phase-quantized block noncoherent channel. Several results, based on the symmetry inherent in the channel model, are provided to characterize this transition density. Low-complexity procedures for computing the channel capacity, and for block demodulation, are obtained using these results. Numerical computations are performed to compare the performance of quantized and unquantized systems, for different quantization precisions, and different block lengths. It is observed, for example, that with QPSK modulation, 8-bin phase quantization of the received signal recovers about 80-85\% of the capacity attained with unquantized observations, while 12-bin phase quantization recovers more than 90\% of the unquantized capacity. Dithering the constellation is shown to improve the performance in the face of drastic quantization.
\end{abstract}


\section{Introduction}
The economies of scale provided by integrated circuit implementation of sophisticated digital signal processing (DSP) algorithms have propelled mass market deployment of cellular and wireless local area network systems over the last two decades. As we now look to scale up the speeds of such DSP-centric architectures by orders of magnitude (e.g., to build multigigabit/sec systems by exploiting the wide swath of spectrum in the 60 GHz band \cite{Smulders_60GHz}), the analog-to-digital converter (ADC), which converts the received analog waveform into the digital domain, becomes a bottleneck: high-speed high-precision ADC is costly and power-hungry \cite{Walden, ADC_survey_2005}. It is of interest, therefore, to explore the feasibility of system design with low-precision ADC. Recent information-theoretic results \cite{TCOM_09, ISIT08} show that for an ideal channel model (perfect synchronization, no dispersion), the loss in the Shannon capacity incurred by using 1-3 bits ADC is very acceptable, even at moderately high signal-to-noise ratio ($\snr$). In this paper, we consider a system without a priori carrier synchronization, i.e., there is a small frequency offset between the receiver's local oscillator and the incoming carrier wave, and investigate the impact of low-precision ADC on the achievable performance.


To model the effect of carrier asynchronism, we consider a discrete-time complex baseband {\it block noncoherent} additive white Gaussian noise (AWGN) channel: if the receiver's local oscillator is not synchronized with that of the transmitter, the phase after downconversion is a priori unknown, but, for practical values of carrier offset, well approximated as constant over a block of symbols. The classical approach to noncoherent communication is to approximate the phase as constant over two symbols, and to apply differential modulation and demodulation. Divsalar and Simon \cite{Divsalar_Simon} were the first to point out the gains that may be achieved by performing multiple symbol differential demodulation over a block of $L>2$ symbols, under the assumption that the phase remains constant over the entire block ($L$ is termed the channel coherence length). More recent work \cite{Warrier, Sweldens, Chen_Tcom} has shown that block demodulation, even for large $L$, can be implemented efficiently, and exhibits excellent performance for both coded and uncoded systems.


We study the effect of low-precision receiver quantization for the block noncoherent channel, under $M$-ary Phase Shift Keying (MPSK) modulation. Since PSK encodes information in the phase of the transmitted symbols, and since the channel impairment is also being modeled as a phase rotation, we investigate an architecture in which the receiver quantizes {\it only the  phase} of the received samples, disregarding the amplitude information. Such phase-only quantization is attractive  because of its ease of implementation: it eliminates the need for automatic gain control (since no amplitude information is used), and can be performed using only $1$-bit ADCs preceded by analog pre-multipliers. For example, 1-bit ADC on both I and Q channels implements uniform 4-sector phase quantization. Uniform 8-sector quantization may simply be achieved by adding two new linear combinations, I+Q and I-Q (no analog pre-multipliers needed in this case), corresponding to a $\frac{\pi}{4}$ rotation of the I/Q axis (see Fig. \ref{fig:Receiver}).

Our focus here is on computing the capacity and the uncoded error rates for communication over the phase-quantized block noncoherent  channel. While brute force computation of these quantities has complexity that scales exponentially in the coherence length of the channel, we find that significant complexity reduction can be attained by understanding the nature of the channel block transition probability. A summary of our analytical and numerical results, for $M$-PSK modulation and uniform $K$-sector phase quantization, is as follows:

\begin{enumerate}
\item  We begin by studying the structure of the input-output relationship of the phase quantized block noncoherent AWGN channel. For the special case when $M$ divides $K$, we exploit the symmetry inherent in the channel model to derive several results characterizing the output probability distribution over a block of symbols, both conditioned on the input, and without conditioning. These results are used to obtain a low-complexity procedure for computing the capacity of the channel (brute force computation has complexity exponential in the block length $L$).
\item We obtain low-complexity optimal block noncoherent demodulation rules. These rules are obtained by specializing the existing low-complexity procedures for block demodulation with unquantized observations, to our setting with quantized observations. 
\item   A close analysis of the block demodulator reveals that, depending on the number of quantization sectors, the symmetries inherent in the channel model, while helping us reduce the computational complexities, can also have a dire consequence: they can make it impossible to distinguish between the effect of the unknown phase offset and the phase modulation. As a result, we may have two equally likely inputs for certain outputs, irrespective of the block length and the $\snr$, leading to severe performance degradation. In order to break the undesirable symmetries, we investigate the performance with a {\it dithered}-PSK input scheme, in which we rotate the PSK constellation across the different symbols in a block. 
\item Numerical results are obtained for QPSK input with 8 and 12 sector phase quantization, for different choices of the block length $L$, and compared with the unquantized performance (studied earlier in \cite{Peleg} as well). We find that 8-sector quantization, with a dithered-QPSK input, achieves more than 80-85 \% of the capacity achieved with unquantized observations (with an identical block length), while with 12-sector quantization, and no dithering, we can get as much as 90-95 \% of the unquantized capacity. The corresponding loss in terms of $\snr$, for fixed capacity, varies between 2 -- 4 dB for 8-sector quantization, and between 0.5 -- 2 dB with 12 sectors. In terms of the uncoded symbol error rates ($\mathsf{SER}$), the performance degradation is of the same order. For instance, at $\mathsf{SER}=10^{-3}$, the loss for 8 and 12 sector quantization, compared to unquantized observations, is about 4 dB and 2 dB respectively.
\end{enumerate}

\noindent {\bf Related work:} 
Two recent works, that are most closely related to our work  are \cite{Sobel_JSSC, Krone_PO}. Both these works consider a phase offset between the transmitter and the receiver, and investigate the impact of output quantization on the achievable performance. In \cite{Sobel_JSSC}, the authors study a mixed-signal receiver architecture, wherein the unknown phase offset is first estimated in post-ADC DSP (in a data aided, or, a non data aided manner), and then compensated for in the analog domain prior to the ADC. In \cite{Krone_PO}, the authors do not resort to analog domain compensation (e.g., due to hardware complexity constraints), but still use the knowledge of the phase offset (assumed to be somehow available) in the post-ADC processing. They observe that, in such a system, output (amplitude) quantization can cause significant degradation in the achievable transmission rate. Note that while both these works focus on \emph{explicit} estimation/knowledge and use of the phase offset, the block noncoherent receiver studied here, rather amounts to an \emph{implicit} joint estimation of the unknown phase offset and the unknown data symbols. Besides these works, and our own results reported earlier in \cite{ISIT09, ICUWB09}, we are not aware of any other work that investigates the impact of output quantization in a carrier asynchronous receiver.

In this work, we consider only a phase offset (induced by the asynchronous local oscillators) between the transmitter and the receiver. However, the block noncoherent model extends to the setting of communication over narrowband slow \emph{fading} channels as well, where the channel state, although unknown, can be assumed to be constant over a block of symbols. Indeed, this block fading model has been investigated extensively in the recent literature, ranging from capacity analysis \cite{hochwald}, to efficient architectures for block demodulation and decoding (\cite{Warrier, Sweldens, Chen_Tcom, jacobsen}, and references therein). 
While our work draws upon this extensive literature, it addresses, for the first time, the unique channel characteristics that arise due to (drastic) output quantization of the block noncoherent channel.

Our prior information-theoretic studies to understand the impact of output quantization on the capacity of the ideal AWGN channel include \cite{SPAWC06, ISIT08, TCOM_09}. More recent work that explores the impact of quantization on other aspects of communication system design includes, amongst others, the study of oversampling \cite{Koch_OS, Krone_OS}, non-symmetric quantization\cite{Koch_ASQ}, automatic gain control \cite{Sun_AGC}, channel estimation performance \cite{Onkar_CE}, performance over fading channels \cite{Krone_Fading}, broadcast and MAC channels \cite{Chandrasekaran}, and channels with memory \cite{Zeitler}, constellation shaping methods \cite{Valenti}, and study of system performance under the generalized mutual information criterion \cite{Zhang}.

\noindent {\bf Organization of the Paper:} The rest of the paper is organized as follows. In Section \ref{Sec:ChannelModel}, we describe the channel model and the receiver architecture for phase quantization. In Section \ref{Sec:Input-Output}, we study the properties of the channel's transition probability distribution. Efficient mechanisms for computing the channel capacity and for block noncoherent demodulation are described in Sections \ref{Sec:Capacity} and \ref{Sec:BlockDemod}, respectively. Numerical results are presented in Section \ref{Sec:Results}, followed by the conclusions and a list of open issues in Section \ref{Sec:Conclusions}.


{\it Notation}: Throughout the paper, we denote random variables by
capital letters, and the specific value they take using small
letters. Bold faced notation is used to denote vectors of random
variables. $\Ex$ is the expectation operator.

\section{Channel Model and Receiver Architecture} \label{Sec:ChannelModel}

The received signal over a block of length $L$, after quantization is represented as
\begin{equation}\label{eq:Channel_Model}
{Z_l} = \mathsf{Q}(S_le^{j\Phi} +N_l) \ , \ l=0,1,\cdots,L-1,
\end{equation}
where,
\begin{itemize}
\item ${\bf S}:=[S_0  \ S_1  \ \cdots  \ S_{L-1}]$ is the transmitted vector,
\item $\Phi$ is an unknown constant with uniform distribution on $[0, 2\pi)$,
\item $\N:=[N_0 \  \cdots \  N_{L-1}]$ is a vector of i.i.d. complex Gaussian noise with variance $\sigma^2=N_0/2$ in each dimension,
\item $\mathsf{Q}:\complex \tends \mathcal{K}=\{0,1,\cdots,K-1\}$ denotes a quantization function that maps each point in the complex plane to one of the $K$ quantization indices, and
\item $\Z:=[Z_0 \  Z_1 \  \cdots \  Z_{L-1}]$ is the vector of quantized received symbols, so that each $Z_l \in \mathcal{K}$.
\end{itemize}
Each $S_l$ is picked in an i.i.d. manner from a uniform M-PSK
constellation denoted by the set of points
$\mathcal{A}=\{e^{j\theta_0}, e^{j\theta_1}, \cdots,
e^{j\theta_{M-1}}\}$, where $\theta_m=(\theta_{m-1}+\frac{2\pi}{M})$
{\footnote{Unless stated otherwise, any arithmetic operations for
phase angles are assumed to be performed modulo $2\pi$. For the output
symbols $Z_l$, the arithmetic is modulo $K$, while for the input
symbols $X_l$ (introduced immediately after in the text ), it is
modulo M.}}, for $m=1,2,\cdots,M-1$.

We now introduce the random vector $\X=[X_0 \ \ X_1 \ \ \cdots \ \ X_{L-1}]$, with each $X_i$ picked in an i.i.d. manner from a uniform distribution on the set $\{0,1,\cdots,M-1\}$. Our channel model \eqref{eq:Channel_Model} can now equivalently be written as
\begin{equation}\label{eq:Channel_Model_e}
{Z_l} = \mathsf{Q}(e^{j\theta_{X_l}}e^{j\Phi} +N_l) \ , \ l=0,1,\cdots,L-1 \ ,
\end{equation}
with every output symbol $Z_l \in \{0,1,\cdots,K-1\}$ as before, and
every input symbol $X_l \in \{0,1,\cdots,M-1\}$. The set of all
possible input vectors is denoted by $\mathcal{X}$, while
$\mathcal{Z}$ denotes the set of all possible output vectors.

\begin{figure}[]
      \centerline { \epsfig{figure=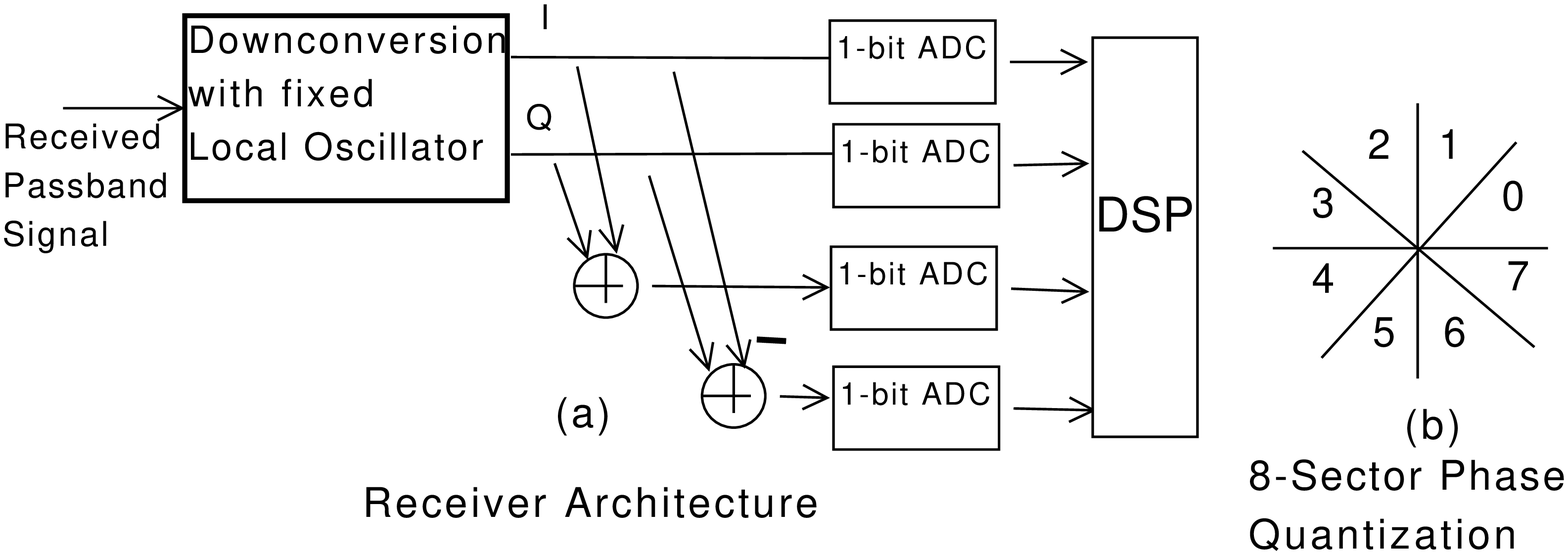, width=14cm, height=5cm}}
      \caption{Receiver architecture for 8-sector quantization.}
       \label{fig:Receiver}
 \end{figure}

We consider $K$-bin (or $K$-sector) phase quantization: our quantizer
divides the interval $[0,2\pi)$ into $K$ equal parts, and the
quantization indices go from $0$ to $K-1$ in the counter-clockwise
direction. Fig. \ref{fig:Receiver}(b) depicts the scenario for $K$=8.
Thus, our quantization function is $\mathsf{Q}(c)=\lfloor
\arg(c)|(\frac{2\pi}{K}) \rfloor$, where $c \in \complex$, and
$\lfloor p \rfloor$ denotes the greatest integer less than or equal to
$p$. Such phase quantization can be implemented using $1$-bit ADCs
preceded by analog multipliers which provide linear combinations of
the $I$ and $Q$ channel samples. For instance, employing $1$-bit ADC on $I$ and $Q$ channels results in uniform $4$-sector phase quantization, while uniform $8$-sector quantization can be achieved simply by adding two new linear combinations, $I$+$Q$ and $I$-$Q$, corresponding to a $\pi/4$ rotation of $I$/$Q$ axes (no analog multipliers needed in this case), as shown in Fig. \ref{fig:Receiver}(a).\looseness-1


We begin our investigation by studying the inherent symmetry in the relationship between the channel input and output. This yields several results that govern the structure of the output probability distribution, both conditioned on the input (i.e., $\prob(\Z|\X)$), and without conditioning (i.e., $\prob(\Z)$). These distributions are integral to computing the channel capacity (one of our focuses in this paper), as well as for soft decision decoding (not considered here).  While brute force computation (computing $\prob(\z|\x)$ for every $\z\in \mathcal{Z}$ and every $\x \in \mathcal{X}$) of these distributions has exponential complexity in the block length, our results allow their computation with significant reduction in the complexity.

{\it Note:} Throughout the paper, we assume that the PSK constellation size $M$, and the number of quantization bins $K$, are such that $K=aM$ for some positive integer $a$. We illustrate our results with the running example of QPSK with 8-sector quantization (so that $a=2$), depicted in Fig. \ref{fig:Examples}(a)


\section{Input-Output Relationship}\label{Sec:Input-Output}



Conditioned on the channel phase $\Phi$, $\prob(\Z|\X , \Phi)$ is
a product of individual symbol probabilities $\prob(Z_l|X_l,\Phi)$.  We therefore begin by analyzing the symmetries in the latter.

\subsection{Properties of $\prob(Z_l|X_l,\Phi)$}
We have that $\prob(z_l|x_l,\phi)$ is the probability that $\arg(e^{j(\theta_{x_l}+\phi)} +N_l)$ belongs to the interval $[\frac{2\pi}{K}z_l \ \ \frac{2\pi}{K}(z_l+1)).$ In other words, it is the probability that the complex Gaussian noise $N_l$ takes the point $e^{j(\theta_{x_l}+\phi)}$ on the unit circle, to another point whose phase belongs to $[\frac{2\pi}{K}z_l \ \ \frac{2\pi}{K}(z_l+1))$. Due to the circular symmetry of the complex Gaussian noise, this is the same as the probability that $N_l$ takes the point $e^{j(\theta_{x_l}+\phi+\frac{2\pi}{K}i)}$ on the unit circle, to another point whose phase belongs to $[\frac{2\pi}{K}(z_l+i) \ \ \frac{2\pi}{K}(z_l+1+i))$, where $i$ is an integer. We thus get our first two results.

{\it Property A-1:} $\prob(z_l|x_l,\phi)=\prob(z_l+i|x_l,\phi+i\frac{2\pi}{K}).$

{\it Property A-2:} $\prob(z_l|x_l,\phi)=\prob(z_l+ia|x_l+i,\phi).$

Note that $\theta_{x_l+i}=\theta_{x_l}+\frac{2\pi}{M}i= \theta_{x_l}+\frac{2\pi}{K}(ia)$, which gives Property {\it A-2}. 

\begin{figure}[t]
      \centerline { \epsfig{figure=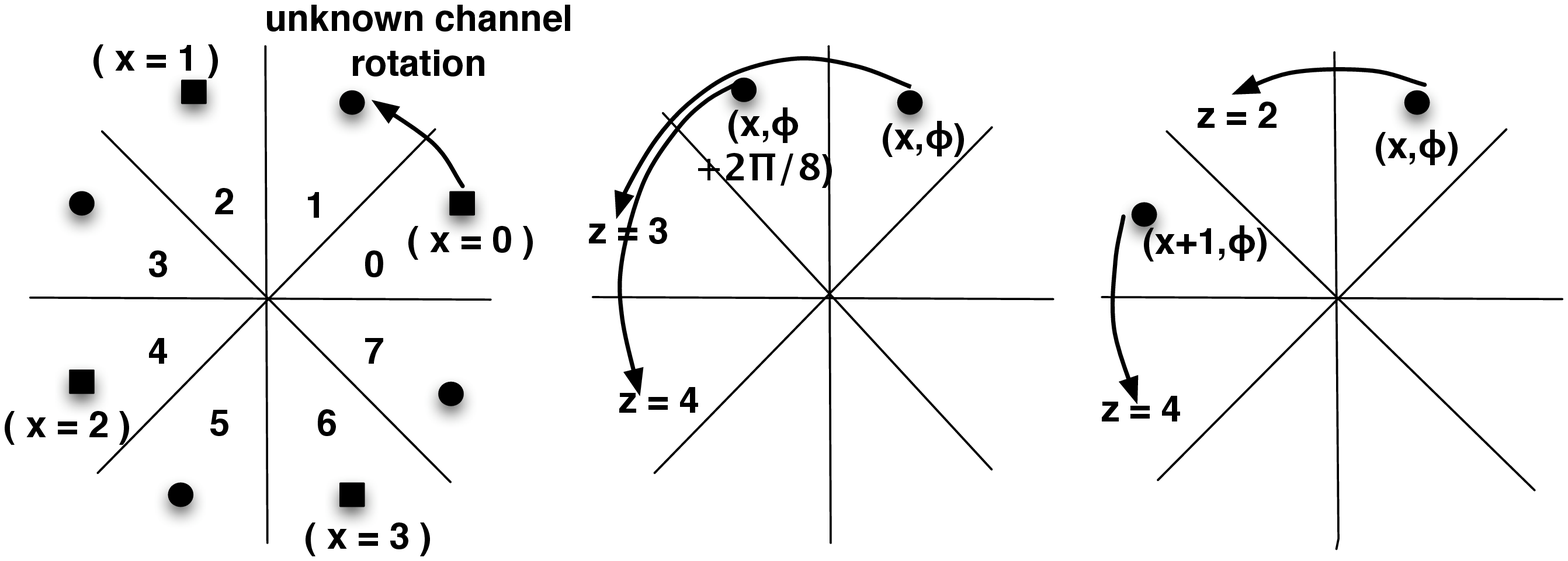, width=13cm, height=5cm}}
      \caption{QPSK with 8-sector quantization (i.e., M=4, K=8). \ a) depicts how the unknown channel phase $\phi$ results in a rotation of the transmitted symbol (square : original constellation , circle : rotated constellation). (b) and (c) depict the circular symmetry induced in the conditional probability $\prob(z|x,\phi)$ due to the circular symmetry of the complex Gaussian noise. \ \ (b) shows that increasing $\phi$ by $2\pi/K=(\pi/4)$  and $z$ by $1$ will keep the conditional probability unchanged, i.e., $\prob(z=3|x,\phi)=\prob(z=4|x,\phi+2\pi/K)$. \ \ (c) shows that increasing $x$ by $1$ and $z$ by $2 =(K/M)$ will keep the conditional probability unchanged, i.e., $\prob(z=2|x,\phi)=\prob(z=4|x+1,\phi)$.
      }
       \label{fig:Examples}
 \end{figure}

Property $A$-$2$ simply states that if we jump from one point in the M-PSK constellation to the next, then we must jump $a=\frac{K}{M}$ quantization sectors in order to keep the conditional probability invariant. 
This is intuitive, since the separation between consecutive points in the input constellation is $2\pi/M$, while each quantization sector covers an angle of $2\pi/K$.
For QPSK with $K=8$, Fig. \ref{fig:Examples}(b) and \ref{fig:Examples}(c) depict example scenarios for the two properties.

If we put $i=-x_l$ in Property {\it A-2}, we get the following special case, which relates the conditioning on a general $x_l$ to the conditioning on $0$.

{\it Property A-3:} $\prob(z_l|x_l,\phi)=\prob(z_l-ax_l|0,\phi).$

To motivate our final property, we consider our example of QPSK with
$K=8$. While we have $8$ distinct quantization sectors, if we look at
Fig. 2(a), the orientation of these $8$
sectors relative to the $4$ constellation points (shown as squares) can be described
by dividing the sectors into $2$
groups : $\{0,2,4,6\}$, and $\{1,3,5,7\}$. For instance, the
positioning of the first sector ($z=0$) w.r.t. $x=0$ is identical to
the positioning of the third sector ($z=2$) w.r.t. $x=1$ (and
similarly $z=4$ w.r.t $x=2$, and $z=6$ w.r.t $x=3$). On the other
hand, the positioning of the second sector ($z=1$) w.r.t. $x=0$ is
identical to the positioning of the fourth sector ($z=3$) w.r.t. $x=1$
(and similarly $z=5$ w.r.t $x=2$, and $z=7$ w.r.t $x=3$). In
terms of the conditional probabilities, this implies, for example,
that we will have $\prob(z_l=7|x_l=3,\phi)=\prob(z_l=1|x_l=0,\phi)$,
and similarly, $\prob(z_l=6|x_l=3,\phi)=\prob(z_l=0|x_l=0,\phi)$.
In general, we can relate the conditional probability of every odd $z_l$ with that of
$z_l=1$, and similarly of every even $z_l$ with that of $z_l=0$,
with corresponding rotations of the symbol $x_l$. For general values of $K$ and $M$, the number of groups equals $a=\frac{K}{M}$, and we can relate the probability of any $z_l$ with that of $z_l \bmod a$.\looseness-1 


{\it Property A-4:} Let $z_l= q_la+r_l$, where $q_l$ is the quotient on dividing $z_l$ by $a$, and $r_l$ is the remainder, i.e, $r_l=z_l \bmod a$. Then, $\prob(z_l|x_l,\phi)=\prob(z_l \bmod a | x_l-q_l ,\phi)$.

While this result follows directly from Property $A$-$2$ by putting
$i=-q_l$, it is an important special case, as it enables us to restrict attention to only the first $a$ sectors ($Z_l \in \{0,1,\cdots,a-1\}$), rather than having to work with all the $K$ sectors. As detailed later, this leads to significant complexity reduction in capacity computation.

We now use these properties to present results for $\prob(\Z|\X)$.
\subsection{Properties of $\prob(\Z|\X)$}

{\emph{Property B-1}}: Let $\pmb{1}$ denote the row vector with all entries as $1$. Then $\prob(\z|\x)=\prob(\z+i\pmb{1}|\x)$.

{\it Proof:} For a fixed $\x$, increasing each $z_l$ by the same number $i$ leaves the conditional probability unchanged, because the phase $\Phi$ in the channel model \eqref{eq:Channel_Model} is uniformly distributed in $[0,2\pi)$.
A detailed proof follows. We have
\begin{equation*}
\begin{split}
\prob(\z|\x)& =\Ex_{\Phi}\left({\prob(\z|\x,\Phi)}\right) = \Ex_{\Phi}\left({\prod_{l=0}^{L-1}{\prob(z_l|x_l,\Phi)}}\right) \\
& = \Ex_{\Phi}\left({\prod_{l=0}^{L-1}{\prob(z_l+i|x_l,\Phi+i\frac{2\pi}{K})}}\right)\\
& = \Ex_{\hat \Phi}\left({\prod_{l=0}^{L-1}{\prob(z_l+i|x_l, \hat{\Phi}}})\right)\\
& = \Ex_{\hat \Phi}\left({\prob(\z+i\pmb{1}|\x,\hat\Phi)})\right) = \prob(\z+i\pmb{1}|\x).
\end{split}
\end{equation*}
The second equality follows by the fact that the components of $\Z$ are independent conditioned on $\X$ and $\Phi$. Property $A$-$1$ gives the third equality. A change of variables, $\hat{\Phi}=\Phi+i\frac{2\pi}{K}$ gives the fourth equality (since $\Phi$ is uniformly distributed on $[0, 2\pi)$, so is $\hat{\Phi}$), thereby completing the proof. \qed  \looseness-1
{\it Remark 1:} For the rest of the paper, we refer to the operation $\z \tends\ \z+i\pmb{1}$ as {\it constant addition}.

Next, we observe that the conditional probability remains invariant under an {\it identical}
permutation of the components of the vectors $\z$ and
$\x$. \looseness-1

{\emph{Property B-2}}: Let $\Pi$ denote a permutation operation, and $\Pi\x \ (\Pi\z)$ the vector obtained on permuting $\x \ (\z)$ under this operation. Then, $\prob(\z|\x)=\prob(\Pi{\z}|\Pi{\x})$.

{\it Proof:} As in the proof of Property $1$, the idea is to condition on $\Phi$ and work with the symbol probabilities $\prob(z_l|x_l,\Phi)$. Consider $\prob(\z|\x,\Phi) ={\prod_{l=0}^{L-1}{\prob(z_l|x_l,\Phi)}}$, and $\prob(\Pi{\z}|\Pi{\x},\Phi)= {\prod_{l=0}^{L-1}{\prob({(\Pi \z)}_l|{(\Pi \x)}_l,\Phi)}}$. Since multiplication is a commutative operation, we have $\prob(\z|\x,\Phi)=\prob(\Pi{\z}|\Pi{\x},\Phi)$. Taking expectation w.r.t. $\Phi$ completes the proof. \qed

The next two results extend properties $A$-$3$ and $A$-$4$.

{\it Property B-3:} Define the input vector $\x_0=[0 \cdots 0]$. Then, $\prob(\z|\x)=\prob(\z-a\x|\x_0)$, where $a=\frac{K}{M}$, and the subtraction is performed modulo $K$.

{\it Property B-4:} Let $z_l= q_la+r_l$, where $q_l$ is the quotient on dividing $z_l$ by $a$, and $r_l$ is the remainder, i.e, $r_l=z_l \bmod a$. Define $\q=[q_0,\cdots,q_{L-1}]$, and, $\z \bmod a =[z_0 \bmod a \ \  \cdots \ \ z_{L-1} \bmod a]$. Then $\prob(\z|\x)=\prob(\z \bmod  a \ | \ \x-\q) $.

{\it Proofs:} The properties follow from $A$-$3$ and $A$-$4$ respectively, by first noting that the vector probability $\prob(\z|\x,\Phi)$ is the product of the scalar ones, and then integrating over $\Phi$ .\qed
\subsection{Properties of $\prob(\Z)$}

We now consider the unconditional distribution $\prob(\z)$. The first
result states that $\prob(\z)$ is invariant under constant addition.

{\it Property C-1:} $\prob(\z)=\prob(\z+i\pmb{1})$.

{\it Proof:} Using Property $B$-$1$, this follows directly by taking expectation over $\X$ on both sides.
\qed

We now extend Property $B$-$2$ along similar lines to show that $\prob(\z)$ is invariant under any permutation of $\z$.

{\emph{Property C-2:}} $\prob(\z)=\prob(\Pi{\z})$.

{\it Proof:} We have $\prob(\z) =\frac{1}{M^L}\sum_{\x \in \mathcal{X}}{\prob(\z|\x)}$. Using Property $B$-$2$, we get $\prob(\z) =\frac{1}{M^L}\sum_{\x \in \mathcal{X}}{\prob\left(\Pi{\z}|\Pi\x\right)}.$ Since the permutation operation results in a one-to-one mapping (every unique choice of $\x \in \mathcal{X}$ results in a unique $\Pi\x \in \mathcal{X}$), we can rewrite the last equation as $\prob(\z) \ = \ \frac{1}{M^L}\sum_{\x \in \mathcal{X}}{\prob(\Pi{\z}|\x}) \ = \ \prob(\Pi{\z}).$ \qed

Our final result extends Property $B$-4.

{\it Property C-3}:  Let $a =\frac{K}{M}$. Then $\prob(\z)=\prob(\z \bmod a)$.

{\it Proof:} Using the same notation as in Property $B$-$4$, we have $\prob(\z|\x)=\prob(\z \bmod  a \ | \ \x-\q) \ $. Noting that the transformation $\x \tends \x-\q$ is a one-to-one mapping, the proof follows on the same lines as the proof of Property $C$-$2$.\qed

{\it Example:} For QPSK with $K=8$ and $L=4$, $\prob(z=[5 \ 7 \ 2 \ 4])=\prob(z=[1 \ 1 \ 0 \ 0])$.

We now apply these results for low-complexity capacity computation.

\section{Capacity Computation}\label{Sec:Capacity}

We wish to compute the mutual information 
\[
I(\X;\Z)= H(\Z)-H(\Z|\X) .
\]
We first discuss computation of the conditional entropy.


\subsection{Conditional Entropy}

We have $H(\Z|\X)=\sum_{\mathcal{X}} H(\Z|\x)\prob(\x)$, where
$H(\Z|\x)=-\sum_{\mathcal{Z}} \prob(\z|\x)\log\prob(\z|\x)$ is the entropy
of the output when the input vector $\X$ takes on the specific value
$\x$. Our main result in this section is that $H(\Z|\x)$ is constant
$\forall \x$.


{\emph{Property D-1}}: $H(\Z|\x)$ is a constant.

{\it Proof:} We show that for any input vector $\x, H(\Z|\x)=H(\Z|\x_0)$, where $\x_0=[0 \cdots 0]$ as defined before. We have
\begin{eqnarray}
H(\Z|\x) =&-\displaystyle&\sum_{\mathcal{Z}} \prob(\z|\x)\log\prob(\z|\x) \nonumber \\
=&-\displaystyle&\sum_{\mathcal{Z}} \prob(\z-a\x|\x_0)\log\prob(\z-a\x|\x_0) \label{eq:Cond_Entropy} \ ,
\end{eqnarray}
where the second equality follows from Property $B$-$3$. Noting that the transformation $\z \tends \z-a\x$ is a one-to-one mapping, we can rewrite \eqref{eq:Cond_Entropy} as
\begin{equation}
\begin{split}
H(\Z|\x)& =-\displaystyle\sum_{\mathcal{Z}} \prob(\z|\x_0)\log\prob(\z|\x_0) = H(\Z|\x_0)
\end{split}
\end{equation}
\qed

Thus, $H(\Z|\X) = H(\Z|\x_0)$, but brute force computation of
$H(\Z|\x_0)$ still has exponential complexity: $\prob(\Z|\x_0)$ must
be computed for each of the $K^L$ possible output vectors
$\Z$. However, we find that it suffices to compute $\prob(\Z|\x_0)$ for a much
smaller set of $\Z$ vectors. \looseness-1


Using Property $B$-$2$, we have $\prob(\z|\x_0)=\prob(\Pi \z|\Pi \x_0)$. Since $\x_0=[0 . . 0]$, any permutation of $\x_0$ gives back $\x_0$. Hence, $\prob(\z|\x_0)=\prob(\Pi{\z}|\x_0)$. Combined with Property $B$-$1$, we thus get that it suffices to compute $\prob(\z|\x_0)$ for a set of vectors $S_{\Z}$ in which no vector can be obtained from another by performing the joint operations of constant addition and permutation. While we do not have a method to obtain the set $S_{\Z}$ exactly, we exploit the preceding properties to obtain, and, restrict attention to, a sub-optimal set $S_{\Z_2}$ having cardinality $C(K+L-2,L-1)$, which is still significantly lower than the exponential figure of $K^L$. For instance, with $K=8$, and $L=\{3,4,5,6,7\}$, the cardinality of $S_{\Z_2}$ is $\{36, 120, 330, 792, 1716\}$, whereas the exponential figure $K^L$ is $\{512, 4096, 32768, 2.6\times 10^5, 2.1\times 10^6\}$. Details regarding the construction of the set $S_{\Z_2}$ are provided in Appendix \ref{sec:Appendix1}.



Once we have constructed the set $S_{\Z_2}$, the probability $\prob(\z|\x_0)$ can be numerically computed for every vector $\z$ in this set. Computation of the entropy $H(\Z|\x_0)$ is then straightforward. See Appendix \ref{sec:Appendix1} for the details.

\subsection{Output Entropy}

The output entropy is $H(\Z)=-\sum_{\mathcal{Z}} \prob(\z)\log\prob(\z)$. Brute force computation requires us to know $\prob(\z) \ \forall \z \in \mathcal{Z}$, which clearly has exponential complexity. However, using Properties $C$-$1$, $C$-$2$ and $C$-$3$, we get that it is sufficient to compute $\prob({\z})$ for a set of vectors $\tilde{S}_\Z$ in which no vector can be obtained from another one by performing the operations of constant addition and permutation, and also, the vector components $\in \{0,1,\cdots,a-1\}$.  This is similar to the situation encountered earlier in the last subsection, except that the vector components there were allowed to be in $\{0,1,\cdots,K-1\}$. To exploit this for further complexity reduction, we can begin by defining the set $\tilde{\mathcal{Z}}$ to be the set of vectors in which the vector components take values in $\{0,1,\cdots,a-1\}$  only. Since $\prob(\z)=\prob(\z \bmod a)$, a moment's thought reveals that each vector in $\tilde{\mathcal{Z}}$ has the same probability as a set of $(\frac{K}{a})^L=M^L$ distinct vectors in $\mathcal{Z}$, and the sets corresponding to different vectors are disjoint. Thus $H(Z)=-\displaystyle\sum_{\mathcal{Z}} \prob(\z)\log\prob(\z) = -M^L \displaystyle\sum_{\tilde{\mathcal{Z}}} \prob(\z)\log\prob(\z)$. To obtain $\{\prob(\z)\}$ for $\z \in \tilde{\mathcal{Z}}$, we can follow exactly the same procedure as described in the last subsection, with $K$ being replaced by $a$. In particular, we need to compute $\prob(\z)$ only for $C(a+L-2,L-1)$ vectors.

{\it Example:} For QPSK with $8$ sectors (so $a=2$), the relevant vectors for block length $2$ are $[0 \ \ 0]$ and $[0 \ \ 1]$.

\subsubsection*{Computation of $\prob({\Z})$}
We now need to compute $\prob(\z)=\frac{1}{M^L}\sum_{\x \in \mathcal{X}}\prob(\z|\x)$ for each of the $C(a+L-2,L-1)$ vectors. A brute force approach is to compute $\prob(\z|\x)$ for each $\x$, but again, has exponential complexity.  We exploit the structure in $\z$ to reduce the number of vectors $\x$ for which we need $\prob(\z|\x)$. Specifically, we have that 
each $z_i \in \{0,1,\cdots,a-1\}$. Since there are only $a$ different types of components in $\z$, for block length $L>a$, some of the components in $\z$ will be repeated. For any $\x$, we can then use Property $B$-$2$ to rearrange the components at those locations for which the components in $\z$ are identical, without changing the conditional probability. For instance, let $z_m=z_n$ for some $m,n$. Then, $\prob(\z|\x)=\prob(\z|\Pi{\x})$, where $\Pi{\x}$ is obtained from $\x$ by rearranging the components at locations $m$ and $n$. To sum up, we can restrict attention to a set of vectors $S_\X$ in which no vector can be obtained from another one by permutations between those locations for which the elements in $\z$ are identical. Construction of this set $S_\X$, and the subsequent computation of $\prob({\Z})$, is discussed in Appendix \ref{sec:Appendix2}. For the example scenario of QPSK with 8-sector quantization (so that $a=2$), and block length $L=8$, the worst case cardinality of the set $S_\X$ is 1225, whereas the exponential figure of $M^L=65536$.


\looseness-1

In the next section, we consider efficient block noncoherent demodulation for the phase-quantized channel. This allows us to evaluate the uncoded symbol error rates.
Numerical results for uncoded performance and channel capacity are subsequently provided in Section \ref{Sec:Results}.

\section{Block Noncoherent Demodulation}\label{Sec:BlockDemod}
We consider the generalized likelihood ratio test (GLRT) for block noncoherent demodulation. This entails a joint maximum likelihood estimation of the unknown block of input symbols and the unknown channel phase. Specifically, given the received vector $\z$, the GLRT estimate for $\x$ is given by \looseness-1
\begin{equation}\label{eq:GLRT}
\hat\x(\z)=\displaystyle \argmax_{\x\in\mathcal{X}}\ \max_{\phi \in [0,2\pi)} \ \prob(\z|\x,\phi) \ .
\end{equation}

Brute force computation of the solution to \eqref{eq:GLRT} has prohibitive complexity, since the cardinality of the input space $\mathcal{X}$ grows exponentially with the block length. For unquantized observations, it is known that the solution can rather be obtained with linear-logarithmic complexity \cite{Sweldens}. The key idea used to obtain this complexity reduction works for quantized observations as well, as we illustrate ahead.

First, we make some observations resulting due to the symmetry of our channel model. As before, we let $a=\frac{K}{M}$, and $\q=[q_0 \cdots q_{L-1}]$, where $q_l$ is the quotient obtained on dividing $z_l$ by $a$.
Using Property $A$-4, we get
\begin{equation}
\hat\x(\z)=\displaystyle \argmax_{\x\in\mathcal{X}}\ \max_{\phi \in [0,2\pi)} \ \prob(\z \bmod a|\x-\q,\phi) \ ,
\end{equation}
which in turn gives
\begin{equation}\label{eq:GLRT_1}
\hat\x(\z)=\hat\x(\z \mod a)+\q(\z) \ ,
\end{equation}
where we have explicitly noted that $\q$ is a function of $\z$.
This result is useful in the sense that the solution for a received vector $\z$ can be easily obtained if the solution for $\z \bmod a$ is known, since computing $\q(\z)$ is a trivial task. Hence, we restrict attention to computing the GLRT solution only for those $\z$ for which the vector components $\in \{0,1,\cdots,a-1\}$. Also observe that $\prob(\z|\x,\phi)=\prob(\z|\x+i,\phi-i\frac{2\pi}{M})$. This implies that the demodulator can not distinguish between two input vectors that are related by the operation of constant addition. This is well known (for unquantized observations), and is the basis for using techniques such as differential modulation.

To obtain a low-complexity solution, the key is to interchange the order of maximization in \eqref{eq:GLRT}. Consider \begin{equation}\label{eq:Inner}
\max_{\phi} \max_{\x\in\mathcal{X}} \ \ \prob(\z|\x,\phi) \ .
\end{equation}
For a fixed $\phi$, the inner maximization over $\x$ is straightforward since it can done in a coherent manner, i.e., on a symbol by symbol basis. For $\phi=0$, let the coherent solution be denoted by ${\bf c}(0)=[c_0(0) \cdots c_{L-1}(0)]$. (We dropped the dependence on $\z$ to simplify notation). Note that this means $c_{l}(0)=\displaystyle \argmax_{x_l\in\{0,\cdots,M-1\}} \prob(z_l|x_l,\phi=0)$. As $\phi$ is increased, the coherent solution ${\bf c}$ will change. However, this will happen only when any of the individual solutions $c_l$ changes. The crucial observation now is that as $\phi$ is varied over $0$ to $\frac{2\pi}{M}$, each of the individual solutions $c_l(\phi)$ changes only once. In other words, for each $l$, there is a {\it crossover angle} $\alpha_l$, such that \looseness-1
\begin{equation}
\begin{split}
c_l(\phi) & = c_l(0) \ , \ \ \text{if} \ \ 0 \leq \phi \leq \alpha_l \\
&= c_l(0)+1 \ , \ \ \text{if} \ \ \alpha_l < \phi < \frac{2\pi}{M} \ .
\end{split}
\end{equation}
The exact crossover angles are easy to obtain as functions of $z_l, K, M$ and the locations of the input constellation points. Now, since we only consider those $\z$ vectors for which every component $\in \{0,\cdots,a-1\}$, there can be at most $a$ distinct crossover angles. Hence, when $\phi$ is varied between $[0,\frac{2\pi}{M})$, the number of distinct coherent solutions to the inner maximization in \eqref{eq:Inner} is at most $a$, and these solutions can be obtained simply by sorting the crossover angles in an ascending order. For each of these (at most) $a$ input vectors, we can now numerically compute the metric $\displaystyle\max_{\phi \in [0,2\pi)}\prob({\z|\x,\phi})$, and pick the one with the largest metric as the GLRT solution. This numerical computation can be done, for example, by fine discretization of the interval $[0,2\pi)$, and computing $\prob({\z|\x,\phi})$ for every $\phi$ in this discrete set. The number of computations (multiplications) required to obtain $\displaystyle\max_{\phi \in [0,2\pi)}\prob({\z|\x,\phi})$ then scales linearly in the block length $L$.

Note that we restricted attention to $\phi \in [0,\frac{2\pi}{M})$ only while performing the inner maximization in \eqref{eq:Inner}. This is because if we go on beyond $\frac{2\pi}{M}$, any new solution we get, say ${\bf c}_1$ will be related to one of the existing solutions, say ${\bf c}_2$, by the operation of constant addition, so that the noncoherent demodulator can not distinguish between ${\bf c_1}$ and ${\bf c_2}$.


\section{Numerical Results}\label{Sec:Results}
\begin{figure}[]
      \centerline { \epsfig{figure=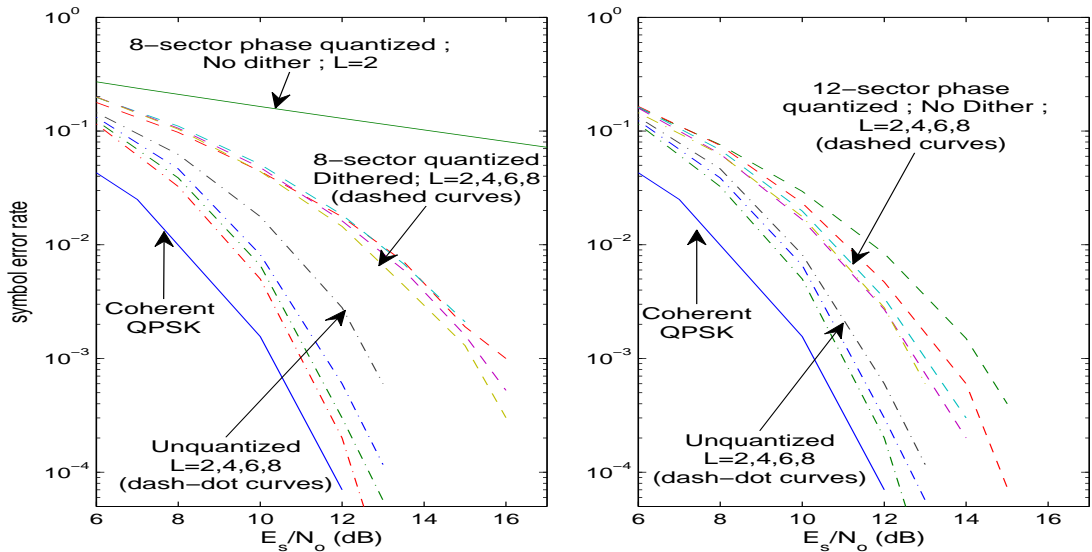, width=17.9cm, height=8cm}}
      \caption{Symbol error rate performance for QPSK with 8-sector phase quantization (left figure) and 12-sector phase quantization (right figure), for block lengths varying from 2 to 8. Also shown for comparison are the curves for coherent QPSK, and noncoherent unquantized QPSK.}
       \label{fig:SER_plots}
\end{figure}

We present results for QPSK with 8-sector and 12-sector phase quantization, for different block lengths L. We begin with the symbol error rate ($\mathsf{SER}$) plots for block demodulation. Fig. \ref{fig:SER_plots} (left plot) shows the results for 8-sector quantization. Looking at the topmost curve, which corresponds to $L=2$, we find that the performance is disastrous. As the $\snr$ is increased, the $\mathsf{SER}$ falls off extremely slowly. A close analysis of the block demodulator reveals that the reason behind this is an ambiguity in the demodulator decision rule: for certain outputs $\z$, irrespective of the $\snr$, the demodulator always returns two equally likely solutions for the input $\x$. While we do not 
\begin{figure}[]
      \centerline { \epsfig{figure=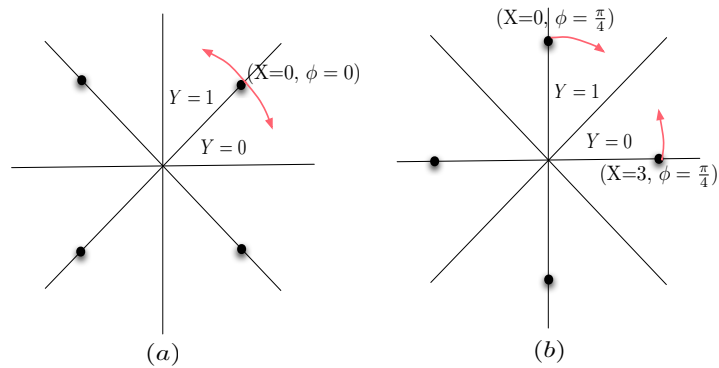, width=9.6cm, height=5cm}}
      \caption{Ambiguity in the block noncoherent demodulator. If the received vector is $[1 \ 0]$, then $(\X=[0 \ 0], \phi=0)$, and, $(\X=[0 \ 3], \phi=\frac{\pi}{4})$ are both equally likely solutions. }
       \label{fig:Ambiguity}
\end{figure}
provide a complete analysis of this ambiguous behavior, an example scenario is shown in Fig. \ref{fig:Ambiguity} to give insight. If the quantized output vector is $\z = [1 \ 0]$, then we find that $\prob(\z|\x,\phi)$ is maximized by two equally likely pairs, $(\x_1,\phi_1)=([0 \ 0], 0)$, and $(\x_2,\phi_2)=([0 \ 3], \pi/4)$, so that the block demodulator, which does joint maximum likelihood estimation over the input and the unknown phase, becomes ambiguous. In other words, the symmetry inherent in the channel model, which on the one hand helped us reduce the complexity of capacity computations, is also making it impossible to distinguish between the effect of the unknown phase offset and the phase modulation on the received signal, resulting in poor performance. While we showed the performance plot for $L=2$ only, we find that the ambiguity persists for larger block lengths also.\looseness-1

\begin{figure}[t]
      \centerline { \epsfig{figure=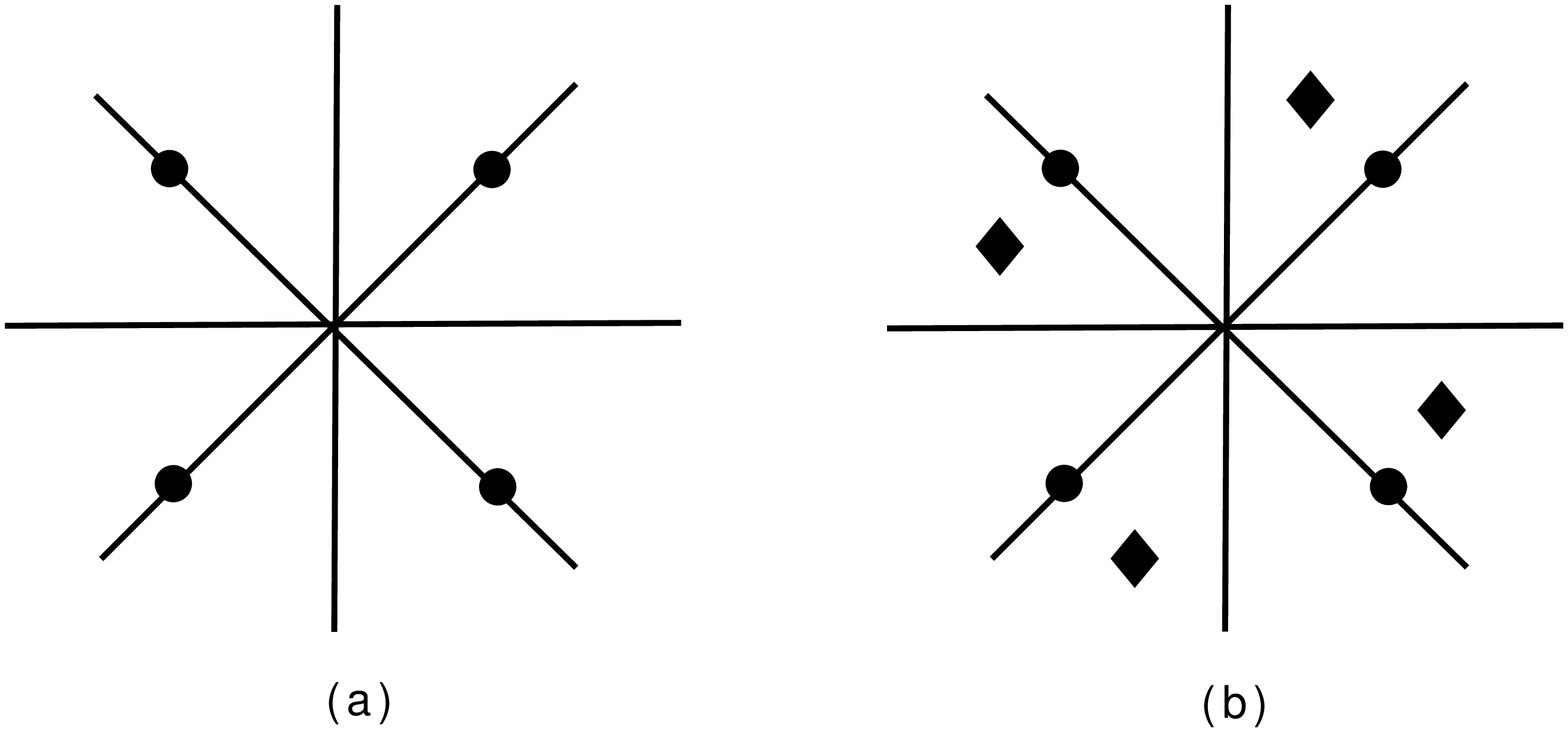, width=8cm, height=3.5cm}}
      \caption{(a) Standard PSK : the same constellation (the one shown) is used for both symbols in the block. (b) Dithered-PSK : the constellations used for the two symbols are not identical, but the second one is a dithered version of the first one.}
       \label{fig:Dither}
\end{figure}

Possible ways to break undesirable symmetries could be to use non-uniform phase quantization, or to employ dithering. Here we investigate the role of the latter. We can either dither the QPSK constellation points at the transmitter, or use analog pre-multipliers to dither the phase quantization boundaries at the receiver. We use a transmit dither scheme in which we rotate the QPSK constellation by an angle of $\frac{1}{L}(\frac{2\pi}{K})$ from one symbol to the next. Fig. \ref{fig:Dither} shows this scheme for block length $L=2$ and $K=8$. The constellation used for the second symbol (shown by the diamond shape) is dithered from the constellation used for the first symbol by an angle of $\pi/8.$ With this choice of transmit constellations, we find that the ambiguity in the block demodulator is removed, and hence the performance is expected to improve. The results in Fig. \ref{fig:SER_plots} (left plot) indeed show a significant performance improvement compared to the no-dithering case, although increasing the block length does not provide much gain. At $\mathsf{SER}$ of $10^{-3}$, 8-sector quantization with $L = 8$ results in a loss of about 4 dB compared to unquantized observations.

On the other hand, if we consider the performance with 12-sector quantization, it is observed that the block demodulator performs well, and dithering is not required. This suggests that 12-sector quantization does not result in any undesirable symmetries in the channel model. Fig. \ref{fig:SER_plots} (right plot) shows the performance for different block lengths. At $\mathsf{SER}$ of $10^{-3}$, and $L=8$, the loss compared to the unquantized observations is reduced to about 2 dB.

\begin{figure}[]
      \centerline { \epsfig{figure=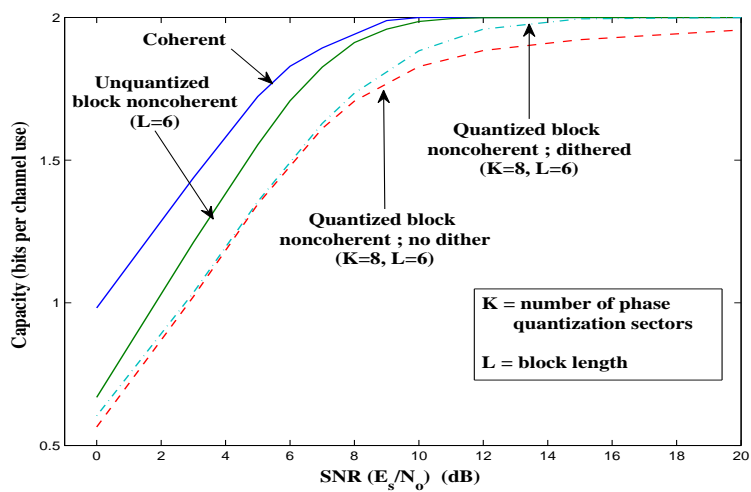, width=11.6cm, height=7cm}}
      \caption{Performance comparison for QPSK with block length $L=6$ : plots depict the capacity of the block noncoherent channel without quantization, and with 8-sector quantization (with and without dithering). Also shown is the capacity for coherent QPSK.}
       \label{fig:Capacity_8_sector}
\end{figure}

Next we show the plots for channel capacity. For all our results, we normalized the mutual information $I(\X;\Z)$ by $L$-1 to obtain the per symbol capacity, since in practice the successive blocks can be overlapped by one symbol due to slow phase variation from one block to the next. Fig. \ref{fig:Capacity_8_sector} shows the results for 8-sector quantization. (To avoid clutter, we show the results for $L=6$ only.) Also shown for reference are the capacity values for the coherent case, and for the block noncoherent case without any quantization. Despite the disastrous performance of the uncoded scheme witnessed earlier, we see that, in terms of the channel capacity, $8$-sector quantization scheme recovers more than $80$-$85$\% of the capacity obtained with unquantized observations, for $\snr >$ 2-3 dB . However, the capacity approaches 2 bits/channel use extremely slowly. Since $H(\X)$ is constant, this implies that $H(\X|\Z)$ falls off very slowly as $\snr \tends \infty$, which is consistent with the earlier observation that there is significant ambiguity in $\X$, given $\Z$, even at high $\snr$. The performance improvement obtained by using a dithered-QPSK input is also shown in Fig. \ref{fig:Capacity_8_sector}. It is seen that the slow increase of capacity towards 2 bits/channel use has been eliminated. {\footnote {Since the low-complexity procedure outlined in Section \ref{Sec:Capacity} does not work once we dither, we used Monte Carlo simulations to compute the capacity with dithering.}}

While the simple transmit dither scheme considered here has improved the performance (in terms of both the $\mathsf{SER}$, as well as channel capacity), we hasten to add that there is no optimality associated with it. A more detailed investigation of different dithering schemes and their potential gains is therefore an important topic for future research.

\begin{figure}[]
      \centerline { \epsfig{figure=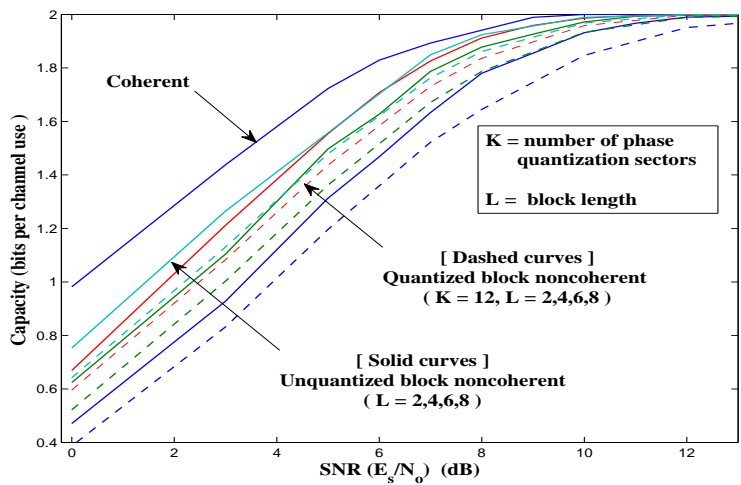, width=11.6cm, height=7cm}}
      \caption{Performance comparison for QPSK : plots depict the capacity of the coherent channel, unquantized block noncoherent channel (different block lengths), and the 12-sector quantized block noncoherent channel (different block lengths).}
       \label{fig:Capacity_12_sector}
\end{figure}

In Fig. \ref{fig:Capacity_12_sector}, we plot the capacity curves for QPSK with 12-sector quantization, for block length $L=\{2,4,6,8\}$. Also shown for reference are the coherent and unquantized block noncoherent performance curves. For identical block lengths, the loss in capacity (at a fixed $\snr >$ 2-3 dB) compared to the unquantized case is less than 5-10 \%, while the loss in power efficiency (for fixed capacity) varies between 0.5-2 dB, and as before, dithering is not required.\looseness-1

\section{Conclusions}\label{Sec:Conclusions}

We have investigated the capacity limits and the uncoded error rates imposed by the use of low-precision phase quantization in a carrier-asynchronous receiver. Building up on our earlier results obtained under ideal channel conditions \cite{TCOM_09}, the results here indicate that low-precision quantization could be a feasible option to overcome the ADC bottleneck in practical high-speed wireless system design. Two critical observations, that we make, however, are that low-precision quantization might lead to unexpected and ambiguous receiver operation when dealing with unknown parameters, and, mechanisms such as dithering might be essential to attain good performance in the face of such ambiguities. Indeed, with low-precision ADC, dithering has been found to be useful while dealing with other aspects of receiver design as well : see \cite{Onkar_CE, Sun_AGC} for the crucial role played by dithering in channel estimation and automatic gain control with quantized receiver observations.




There are several open issues to be addressed. Given the performance improvement obtained using the simple dithering scheme considered here, a more detailed investigation of different dithering schemes is required. Another possibility to consider, which we have not explored here, would be non-uniform phase quantization. While we have restricted attention to PSK inputs in this work, it is important to evaluate performance with QAM alphabets as well, in which case we need to consider amplitude quantization. Note that, amplitude quantization can, in principle help improve performance with PSK inputs as well, especially if the $\snr$ is low, and the block lengths are small. Another topic of interest is the study and development of practical capacity approaching coded modulation strategies for phase quantized communication. An important practical issue is determining whether timing synchronization (which is assumed in the model here) can also be attained using phase-quantized samples, or whether some form of additional information (perhaps using analog techniques prior to the ADC) is required.

As with prior work in the literature, we assumed that the phase across the different blocks varies independently. While this allows analytical tractability, the continuous variation of the phase from one block to the next can be used to enhance performance, especially when we are constrained to using low-precision samples. How best to leverage this memory might be worth investigating.


\bibliographystyle{IEEEtran}
\bibliography{References}

\appendices
\section{Computation of $H(\Z|\x_0)$}\label{sec:Appendix1}
Instead of jointly accounting for constant addition and permutation, we first account for constant addition, and then for permutation. Specifically, we first note that using Property $B$-$1$, it suffices to compute $\prob(\z|\x_0)$ only for a set of vectors $S_{\Z_1}$ for which the first symbol is $0$. Next, using the fact that $\prob(\z|\x_0)=\prob(\Pi{\z}|\x_0)$, within the set $S_{\Z_1}$, we can further restrict attention to a subset $S_{\Z_2}$ in which no vector can be obtained from another one by a permutation operation. Since permutations don't matter, all we are interested in is how many symbols of each type are picked, so that obtaining the set $S_{\Z_2}$ is equivalent to the well-known problem of distributing $L$--$1$ identical balls into $K$ distinct boxes, with empty boxes allowed. The number of ways to do this is $C(K+L-2,L-1)$, and each of these combinations can be obtained easily using standard known procedures.

Once we have the set $S_{\Z_2}$, we can numerically compute the probability
$\prob(\z|\x_0)$ for every vector in $S_{\Z_2}$. The entropy $H(\Z|\x_0)$ can then be obtained as follows. For $\z \in S_{\Z_2}$, let $n(\z)$ denote the number of distinct permutations that can be generated from it, while keeping the first symbol fixed. This is equal to $\frac{(L-1)!}{\prod_{i=0}^{K-1}{r_i}}$, where $r_i$ is the number of times the symbol $i$ occurs in $\z$. The conditional entropy then is $H(\Z|\x_0)=
-\displaystyle\sum_{\mathcal{Z}} \prob(\z|\x_0)\log\prob(\z|\x_0) = -\displaystyle\sum_{S_{\Z_1}} K \prob(\z|\x_0)\log\prob(\z|\x_0) =-\displaystyle\sum_{S_{\Z_2}} K \ n(\z) \ \prob(\z|\x_0)\log\prob(\z|\x_0)$. 

\section{Computation of $P(\Z)$}\label{sec:Appendix2}
To obtain the set $S_\X$, we divide the $L$ locations into $a$ groups, and permutations are allowed only between locations belonging to the same group. The problem then breaks down into $a$ sub-problems. Specifically, let the number of locations in the groups be $n_0,n_1,\cdots,n_{a-1}$, then we need to distribute $n_i$ identical balls into $M$ distinct boxes, for each $i$. The required number of combinations is the product of the individual solutions. While for large $a$, the reduction in complexity may not be huge, for small values of $a$ (which is the paradigm of interest in this work), the savings will be significant. For instance, for QPSK with $L=8$, and $a=2$, the worst case (which happens when $n_0=n_1$) number of combinations is $1225$, compared to the exponential figure of $M^L=65536$. Once the set $S_X$ has been obtained, we can get $\prob(\z)=\frac{1}{M^L}\sum_{\x \in S_X}q(x)\prob(\z|\x)$. Here, $q(x)=\prod_{i=0}^{a-1}\frac{(n_i)!}{\prod_{j=0}^{M-1}(r_{i,j}(x))!}$ \ , where $r_{i,j}(x)$ is the number of times the input symbol $j$ occurs in the locations belonging to group $i$, for the vector $x$.

\end{document}